\documentstyle[12pt]{article}

\newcommand{\be}{\begin{equation}}
\newcommand{\ee}{\end{equation}}
\newcommand{\bea}{\begin{eqnarray}}
\newcommand{\eea}{\end{eqnarray}}

\begin{document}
\thispagestyle{empty}
\begin{flushright}
ENSLAPP-L-515/95\\
JINR-E2-95-156 \\
hep-th/9504070\\
April 1995
\end{flushright}
\centerline{\large\bf Off-shell (4,4) supersymmetric sigma models
with torsion }

\vspace{0.2cm}
\centerline{\large\bf as gauge theories in harmonic superspace}
\vspace{1truecm}
\vskip0.5cm
\centerline{{\large Evgenyi A. Ivanov}}
\vskip.5cm
\centerline{\it Laboratoire de Physique Th\'eorique, ENSLAPP,}
\centerline{\it ENS Lyon, 46 All\'ee d'Italie, F-69364 Lyon Cedex 07, France}
\centerline{\it and}
\centerline{\it Bogoliubov Laboratory of Theoretical Physics, JINR,}
\centerline{\it 141 980, Dubna near Moscow, Russian Federation}
\vskip.5cm
\vskip1.5cm
\centerline{\bf Abstract}

\vspace{0.3cm}
Starting from the action of $(4,4)$ $2D$ twisted multiplets
in the harmonic superspace with a double set of $SU(2)$ harmonic
variables, we present its generalization which provides
an off-shell description of a wide class of $(4,4)$ sigma models
with torsion and non-commuting left and right complex structures.
The distinguishing features of the action constructed are: (i) a
nonabelian and in general nonlinear gauge
invariance ensuring a correct number of physical degrees of freedom;
(ii) an infinite tower of auxiliary fields.
For a particular class of such models we explicitly demonstrate the
non-commutativity of complex structures on the bosonic target.
\vskip.5cm
\newpage\setcounter{page}1

\noindent{\bf 1. Introduction.}
Remarkable target geometries of $2D$ sigma models with
extended worldsheet SUSY are
revealed most clearly within manifestly supersymmetric off-shell superfield
formulations of these theories. For torsionless $(2,2)$ and $(4,4)$
sigma models the relevant superfield Lagrangians were found to coincide
with (or to be directly related to) the
fundamental objects underlying the given geometry:
K\"ahler potential in the $(2,2)$ case \cite{Zum},
hyper-K\"ahler or quaternionic-K\"ahler potentials in the flat or
curved $(4,4)$ cases [2 - 5].
To have superfield off-shell formulations is also highly
desirable while quantizing these models and proving, e.g., their
ultraviolet finiteness.

An important wide class of $2D$ supersymmetric sigma models
is presented by $(2,2)$ and $(4,4)$ models with torsionful bosonic target
manifolds and two
independent left and right sets of complex structures (see,
e.g. \cite{{HoPa},{GHR}}).
These models and,
in particular, their group manifold WZNW representatives \cite{belg}
can provide non-trivial
backgrounds for
$4D$ superstrings (see, e.g., \cite{Luest}) and be relevant to
$2D$ black holes \cite{RSS}.
A manifestly supersymmetric formulation of $(2,2)$ models with
commuting left and right complex structures in terms of chiral and
twisted chiral $(2,2)$
superfields and an exhaustive discussion of their geometry have been given in
\cite{GHR}. For $(4,4)$ models with commuting structures there exist
manifestly supersymmetric off-shell formulations
in the projective, ordinary and $SU(2)\times SU(2)$ harmonic $(4,4)$
superspaces [10-12]. The appropriate superfields represent, in one or another
way, the $(4,4)$ $2D$ twisted
multiplet \cite{{IK},{GHR}}.

Much less is known about $(2,2)$ and $(4,4)$ sigma models with non-commuting
complex structures, despite the fact that most of the
corresponding group manifold WZNW sigma models \cite{belg} fall into this
category
\cite{RSS}. In particular, it is unclear how
to describe them off shell in general. As was argued in Refs.
\cite{{GHR},{BRL},{RSS}}, twisted $(2,2)$ and $(4,4)$ multiplets are
not suitable for this purpose. It has been then suggested to make use of some
other off-shell representations of $(2,2)$ \cite{{BRL},{DS}} and $(4,4)$
\cite{{BRL},{RIL}} worldsheet SUSY. However, it is an open
question whether the relevant actions correspond to generic sigma models of
this type.

In the present letter we propose another approach to the off-shell
description of general $(4,4)$ sigma models with torsion, based upon an
analogy with general torsionless hyper-K\"ahler $(4,4)$
sigma models in $SU(2)$ harmonic superspace [2 - 4].
We start from
a dual form of the general action of $(4,4)$ twisted superfields in
$SU(2)\times SU(2)$ analytic harmonic superspace with two independent
sets of harmonic variables \cite{IS} and construct a direct
$SU(2)\times SU(2)$ harmonic analog of the hyper-K\"ahler $(4,4)$ action.
The form of the action obtained, contrary to the
torsionless case, proves to be severely constrained by the integrability
conditions following from the commutativity of the left and right harmonic
derivatives. While for four-dimensional bosonic manifolds
the resulting action is reduced to that of twisted superfield,
for manifolds of dimension $4n,\;n\geq 2$, the generic action
{\it cannot be written only in terms of twisted superfields}.
Its most characteristic features are (i) the unavoidable presence of
infinite number of
auxiliary fields and (ii) a nonabelain and in general nonlinear gauge
symmetry which ensures the necessary
number of propagating fields. These symmetry and action
are harmonic analogs of the Poisson gauge symmetry and
actions proposed in \cite{{Ikeda},{austr}}.
For an interesting subclass of these actions, harmonic analogs of
the Yang-Mills ones, we explicitly
demonstrate that the left and right complex
structures on the bosonic target {\it do not commute}.

\vspace{0.3cm}
\noindent{\bf 2. Sigma models in SU(2)xSU(2) harmonic superspace.}
The $SU(2)\times SU(2)$ harmonic superspace is an extension of the
standard real $(4,4)$ $2D$ superspace by two independent sets of harmonic
variables  $u^{\pm 1\;i}$ and $v^{\pm 1\;a}$
($u^{1\;i}u^{-1}_{i} =
v^{1\;a}v^{-1}_{a} = 1$) associated with
the automorphism groups $SU(2)_L$ and $SU(2)_R$ of the left and
right sectors of $(4,4)$ supersymmetry \cite{IS}. The corresponding
analytic subspace is spanned by the following set of coordinates
\be  \label{anal2harm}
(\zeta, u,v) =
(\;x^{++}, x^{--}, \theta^{1,0\;\underline{i}},
\theta^{0,1\;\underline{a}}, u^{\pm1\;i}, v^{\pm1\;a}\;)\;,
\ee
where we omitted the light-cone indices of odd coordinates. The
superscript ``$n,m$'' stands for two independent harmonic $U(1)$ charges,
left ($n$) and right ($m$) ones.

It was argued in \cite{IS} that this type of harmonic superspace is most
appropriate for constructing off-shell formulations of $(4,4)$
sigma models with torsion. This hope mainly relied upon the fact
that the twisted $(4,4)$ multiplet has a natural description as a
real analytic $SU(2)\times SU(2)$
harmonic superfield $q^{1,1}(\zeta,u,v)$ (subjected to some
harmonic constraints). The
most general off-shell action of $n$ such multiplets is
given by the following integral over the analytic superspace
(\ref{anal2harm}) \cite{IS}
\be
S_{q,\omega} = \int \mu^{-2,-2} \{
q^{1,1\;M}(\;D^{2,0} \omega^{-1,1\;M}   +
D^{0,2}\omega^{1,-1\;M} \;) + h^{2,2} (q^{1,1}, u, v) \}\;\; (M=1,...,n) \;.
\label{dualq}
\ee
where
\bea
D^{2,0} &=& \partial^{2,0} + i\theta^{1,0\;\underline{i}}
\theta^{1,0}_{\underline{i}}\partial_{++}\;,
\;\;D^{0,2} \;=\;\partial^{0,2} + i\theta^{0,1\;\underline{a}}
\theta^{0,1}_{\underline{a}}\partial_{--}
\label{harm2der} \\
(\partial^{2,0} &=& u^{1 \;i}\frac{\partial} {\partial u^{-1 \;i}}\;,\;\;
\partial^{0,2} \;=\; v^{1 \;a}\frac{\partial} {\partial v^{-1 \;a}})
\nonumber
\eea
are the left and right analyticity-preserving harmonic derivatives and
$\mu^{-2,-2}$ is the analytic superspace integration measure. In
(\ref{dualq}) the involved superfields are unconstrained analytic,
so from the beginning the action (\ref{dualq}) contains an infinite number
of auxiliary fields coming from the double harmonic expansions with
respect to the
harmonics $u^{\pm1\;i}, v^{\pm1\;a}$. However, after varying
with respect to the Lagrange
multipliers $\omega^{1,-1\;M}, \omega^{-1,1\;M}$, one comes to the action
written only in terms of $q^{1,1\;N}$ subjected to the harmonic constraints
\be  \label{qconstr}
D^{2,0} q^{1,1\;M} = D^{0,2} q^{1,1\;M} = 0\;.
\ee
For each value of $M$ these constraints define the $(4,4)$ twisted
multiplet in the $SU(2)\times
SU(2)$ harmonic superspace ($8+8$ components off-shell), so the action
(\ref{dualq}) is a dual form of the general off-shell action of $(4,4)$
twisted multiplets \cite{IS}.

The crucial feature of the action (\ref{dualq}) is the abelian gauge
invariance
\be
\delta \;\omega^{1,-1\;M} = D^{2,0} \sigma^{-1,-1\;M}
\;, \; \delta \;\omega^{-1,1\;M}
= - D^{0,2} \sigma^{-1,-1\;M}
\label{gauge}
\ee
where $\sigma^{-1,-1\;M}$ are iunconstrained analytic
superfield parameters.
This gauge freedom ensures the on-shell equivalence
of the $q,\omega$ formulation of
the twisted multiplet action to its original $q$ formulation \cite{IS}: it
neutralizes superfluous physical dimension component fields in the
superfields
$\omega^{1,-1\;M}$ and $\omega^{-1,1\;M}$ and thus equalizes the
number of propagating fields in both formulations. It
holds already at the free level, with
$h^{2,2}$ quadratic in $q^{1,1\;M}$, so it is natural to expect that
any reasonable generalization of the action (\ref{dualq}) respects
this symmetry or a generalization of it. We will see soon that this is
indeed so.

The dual twisted multiplet action (\ref{dualq}) is
a good starting point for constructing more general actions which, as we
will show, encompass sigma models with non-commuting left and right
complex structures.

It is useful to apply to the suggestive analogy with the general
action of hyper-K\"ahler
$(4,4)$ sigma models in the $SU(2)$ harmonic superspace \cite{GIKOS}.
This action in the $\omega, L^{+2}$ representation \cite{GIOSap}
looks very similar to (\ref{dualq}), the $SU(2)$ analytic superfield
pair $\omega^M, L^{+2\;M}$ being the clear analog of the $SU(2)\times SU(2)$
analytic superfield triple $\omega^{1,-1\;M}, \omega^{-1,1\;M},
q^{1,1\;M}$ and the general hyper-K\"ahler potential being analogous to
$h^{2,2}$.
However, this analogy breaks in that the hyper-K\"ahler potential is
in general an arbitrary function of all involved
superfields and harmonics while $h^{2,2}$ in (\ref{dualq})
depends only on $q^{1,1\;M}$ and harmonics. Thus an obvious way to
generalize (\ref{dualq}) to cover a wider class of torsionful
$(4,4)$ models is to allow for a dependence on
$\omega^{1,-1\;M}, \omega^{-1,1\;M}$ in $h^{2,2}$.

With these reasonings in mind, we take as an ansatz for the general
action the following one
\be
S_{gen} = \int \mu^{-2,-2} \{
q^{1,1\;M}(\;D^{2,0} \omega^{-1,1\;M}   +
D^{0,2}\omega^{1,-1\;M} \;) + H^{2,2} (q^{1,1}, \omega^{1,-1}, \omega^{-1,1},
u, v) \}\;,
\label{genact}
\ee
where for the moment the $\omega$ dependence in $H^{2,2}$ is not fixed. Now
we are approaching the most important point. Namely, we are going to
show that,
contrary to the case of $SU(2)$ harmonic action of torsionless $(4,4)$
sigma models, the $\omega$ dependence of the potential $H^{2,2}$ in
(\ref{genact}) is completely specified
by the integrability conditions following from the
commutativity relation
\be \label{comm}
[\;D^{2,0}, D^{0,2}\;] = 0\;.
\ee

To this end, let us write the equations of motion corresponding to
(\ref{genact})
\bea
D^{2,0}\omega^{-1,1\;M} + D^{0,2}\omega^{1,-1\;M} &=& -
\frac{\partial H^{2,2} (q,\omega,u,v)}{\partial q^{1,1\;M}}\;,
\label{eqom} \\
D^{2,0}q^{1,1\;M} \;=\;
\frac{\partial H^{2,2} (q,\omega,u,v)}{\partial \omega^{-1,1\;M}}\;, \;\;
D^{0,2}q^{1,1\;M} &=&
\frac{\partial H^{2,2} (q,\omega,u,v)}{\partial \omega^{1,-1\;M}}\;.
\label{eqqu}
\eea
Applying the intgrability condition (\ref{comm}) to the pair of
equations (\ref{eqqu}) and imposing a natural requirement that it
is satisfied as a consequence of the equations of motion (i.e. does not
give rise to any new dynamical restrictions), after some algebra
we arrive at the following set of self-consistency relations
\bea
&& \frac{\partial^2 H^{2,2}}{\partial \omega^{-1,1\;N} \partial
\omega^{-1,1\;M}}
\;=\;
\frac{\partial^2 H^{2,2}}{\partial \omega^{1,-1\;N} \partial
\omega^{1,-1\;M}}
\;=\; \frac{\partial^2 H^{2,2}}{\partial \omega^{1,-1\;(N}
\partial \omega^{-1,1\;M)}}
\;=\; 0\;, \label{consN1} \\
&& \left( \partial^{2,0} + \frac{\partial H^{2,2}}{\partial
\omega^{-1,1\;N}}
\;\frac{\partial}{\partial q^{1,1\;N}}
-{1\over 2}\; \frac{\partial H^{2,2}}{\partial q^{1,1\;N}}
\;\frac{\partial}{\partial \omega^{-1,1\;N}}
\right)
\frac{\partial H^{2,2}}{\partial \omega^{1,-1\;M}} \nonumber \\
&& -\left( \partial^{0,2} + \frac{\partial H^{2,2}}{\partial
\omega^{1,-1\;N}}
\;\frac{\partial}{\partial q^{1,1\;N}}
-{1\over 2}\; \frac{\partial H^{2,2}}{\partial q^{1,1\;N}}
\;\frac{\partial}{\partial \omega^{1,-1\;N}}
\right)
\frac{\partial H^{2,2}}{\partial \omega^{-1,1\;M}} \;=\; 0\;.
\label{consN2}
\eea
Eqs. (\ref{consN1}) imply
\bea
H^{2,2} &=& h^{2,2}(q,u,v) + \omega^{1,-1\;N} h^{1,3\;N}(q,u,v)
+ \omega^{-1,1\;N} h^{3,1\;N}(q,u,v) \nonumber \\
&& + \;\omega^{-1,1\;N}\omega^{1,-1\;M}
h^{2,2\;[N,M]}(q,u,v)\;. \label{Hgen}
\eea
Plugging this expression into the constraint (\ref{consN2}),
we finally deduce four independent constraints on the potentials
$h^{2,2}$, $h^{1,3\;N}$, $h^{3,1\;N}$ and $h^{2,2\;[N,M]}$
\bea
&& \nabla^{2,0} h^{1,3\;N} - \nabla^{0,2} h^{3,1\;N} + h^{2,2\;[N,M]}
\;\frac{\partial h^{2,2}}{\partial q^{1,1\;M}} \;=\; 0 \label{1} \\
&& \nabla^{2,0} h^{2,2\;[N,M]} -
\frac{\partial h^{3,1\;N}}{\partial q^{1,1\;T}} \;h^{2,2\;[T,M]} +
\frac{\partial h^{3,1\;M}}{\partial q^{1,1\;T}}\; h^{2,2\;[T,N]} \;=\; 0
\label{2} \\
&& \nabla^{0,2} h^{2,2\;[N,M]} -
\frac{\partial h^{1,3\;N}}{\partial q^{1,1\;T}}\; h^{2,2\;[T,M]} +
\frac{\partial h^{1,3\;M}}{\partial q^{1,1\;T}}\; h^{2,2\;[T,N]} \;=\; 0
\label{3} \\
&& h^{2,2\;[N,T]}\;\frac{\partial h^{2,2\;[M,L]}}{\partial q^{1,1\;T}} +
h^{2,2\;[L,T]}\;\frac{\partial h^{2,2\;[N,M]}}{\partial q^{1,1\;T}} +
h^{2,2\;[M,T]}\;\frac{\partial h^{2,2\;[L,N]}}{\partial q^{1,1\;T}} \;=\; 0
\label{4}
\eea
where
\be
\nabla^{2,0} = \partial^{2,0} + h^{3,1\;N}\frac{\partial}{\partial
q^{1,1\;N}}
\;,\;\;
\nabla^{0,2} = \partial^{0,2} + h^{1,3\;N}\frac{\partial}{\partial
q^{1,1\;N}}
\;.
\ee
and $\partial^{2,0}, \partial^{0,2}$ act only on the ``target''
harmonics, i.e. those appearing
explicitly in the potentials.

Thus we have shown that the direct generalization of the generic
hyper-K\"ahler $(4,4)$ sigma model action to the torsionful
case is given by the action
\bea
S_{q,\omega} &=&
\int \mu^{-2,-2} \{\; q^{1,1\;M}D^{0,2}\omega^{1,-1\;M} +
q^{1,1\;M}D^{2,0}\omega^{-1,1\;M} +  \omega^{1,-1\;M}h^{1,3\;M}
\nonumber \\
&&+ \omega^{-1,1\;M}h^{3,1\;M} + \omega^{-1,1\;M} \omega^{1,-1\;N}
\;h^{2,2\;[M,N]} + h^{2,2}\;\}\;, \label{haction}
\eea
where the involved potentials depend only on $q^{1,1\;M}$ and target
harmonics and
satisfy the target space constraints (\ref{1}) - (\ref{4}). These
constraints certainly encode a nontrivial geometry which for the time
being is
unclear to us. To reveal it we need to solve the
constraints, which is still to be done. At present we are only aware of
their particular solution which will be discussed in the next
section.

In the rest of this section we present a set of invariances of
the action (\ref{haction}) and constraints (\ref{1}) - (\ref{4})
which can be useful for understanding the underlying geometry of the
given class of sigma models.

One of these invariances is a mixture of reparametrizations in the target
space (spanned by the involved superfields and target harmonics)
and the transformations which are bi-harmonic analogs of hyper-K\"ahler
transformations of Refs. \cite{{BGIO},{GIOSap}}.
It is realized by
\bea
\delta q^{1,1\;N} &=& \lambda^{1,1\;N}\;,\;\; \delta \omega^{-1,1\;N}
\;=\;
-\frac{\partial \lambda^{0,2}}{\partial q^{1,1\;N}} -
\frac{\partial \lambda^{1,1\;M}}{\partial q^{1,1\;N}}\;\omega^{-1,1\;M}\;,
\nonumber \\
\delta \omega^{1,-1\;N} &=&
-\frac{\partial \lambda^{2,0}}{\partial q^{1,1\;N}} -
\frac{\partial \lambda^{1,1\;M}}{\partial q^{1,1\;N}}\;\omega^{1,-1\;M}\;,
\nonumber \\
\delta h^{2,2} &=& \nabla^{2,0} \lambda^{0,2} + \nabla^{0,2}
\lambda^{2,0}\;,
\nonumber \\
\delta h^{3,1\;M} &=& \nabla^{2,0}\lambda^{1,1\;M} + h^{2,2\;[M,N]}\;
\frac{\partial \lambda^{2,0}}{\partial q^{1,1\;N}} \nonumber \\
\delta h^{1,3\;M} &=& \nabla^{0,2}\lambda^{1,1\;M} - h^{2,2\;[M,N]}\;
\frac{\partial \lambda^{0,2}}{\partial q^{1,1\;N}} \nonumber \\
\delta h^{2,2\;[N,M]} &=&
\frac{\partial \lambda^{1,1\;N}}{\partial q^{1,1\;L}}\; h^{2,2\;[L,M]} -
\frac{\partial \lambda^{1,1\;M}}{\partial q^{1,1\;L}}\; h^{2,2\;[L,N]} \;,
\label{hrep}
\eea
all the involved transformation parameters being unconstrained
functions of $(q^{1,1\;M}, u, v)$. This kind of invariance can be used to
bring the potentials in (\ref{haction}) into a ``normal'' form similar to
the normal gauge of the hyper-K\"ahler potential (see \cite{GIOSap}).

Much more interesting is another invariance which has no analog in the
hyper-K\"ahler case and is a nonabelain and in general nonlinear
generalization of the abelian gauge invariance (\ref{gauge})
\bea
\delta \omega^{1,-1\;M}  &=&
\left( D^{2,0}\delta^{MN} + \frac{\partial h^{3,1\;N}}{\partial q^{1,1\;M}}
\right) \sigma^{-1,-1\;N} - \omega^{1,-1\;L} \;
\frac{\partial h^{2,2\;[L,N]}}{\partial q^{1,1\;M}}\;\sigma^{-1,-1\;N}\;,
\nonumber \\
\delta \omega^{-1,1\;M}  &=&
- \left( D^{0,2}\delta^{MN} + \frac{\partial h^{1,3\;N}}{\partial
q^{1,1\;M}} \right) \sigma^{-1,-1\;N} - \omega^{-1,1\;L} \;
\frac{\partial h^{2,2\;[L,N]}}{\partial q^{1,1\;M}}\;\sigma^{-1,-1\;N}\;,
\nonumber \\
\delta q^{1,1\;M} &=& \sigma^{-1,-1\;N} h^{2,2\;[N,M]}\;. \label{gaugenab}
\eea
As expected, the action is invariant only with taking account of the
integrability
conditions (\ref{1}) - (\ref{4}). In general, these gauge
transformations close with a field-dependent Lie bracket parameter.
Indeed, commuting two
such transformations, say, on $q^{1,1\;N}$, and using the
cyclic constraint (\ref{4}), we find
\be
\delta_{br} q^{1,1\;M} = \sigma^{-1,-1\;N}_{br} h^{2,2\;[N,M]}\;, \;\;
\sigma^{-1,-1\;N}_{br} = -\sigma^{-1,-1\;L}_1 \sigma^{-1,-1\;T}_2
\frac{\partial h^{2,2\;[L,T]}}{\partial q^{1,1\;N}}\;.
\ee
We see that eq. (\ref{4}) guarantees the nonlinear closure of the
algebra of gauge transformations (\ref{gaugenab}) and so it is a group
condition similar to the Jacobi identities.

Curiously enough, the
gauge transformations (\ref{gaugenab}) augmented with the group
condition (\ref{4})
are precise bi-harmonic counterparts of the two-dimensional version of
basic relations of
the Poisson nonlinear gauge theory
which received some attention recently
\cite{{Ikeda},{austr}} (with the evident correspondence
$D^{2,0}, D^{0,2} \leftrightarrow \partial_\mu$;
$\omega^{1,-1\;M}, - \omega^{-1,1\;M} \leftrightarrow
A_\mu^M $; $\mu = 1,2$). The action (\ref{haction})
coincides in appearance with the general (non-topological) action of
Poisson gauge theory \cite{austr}. The manifold $(q,u,v)$ can be
interpreted as a kind of bi-harmonic extension of some Poisson
manifold and the potential
$h^{2,2\;[N,M]}(q,u,v)$ as a tensor field inducing the Poisson structure on
this extension. We find it remarkable that the harmonic superspace action of
torsionful $(4,4)$ sigma models deduced using an analogy with
hyper-K\"ahler $(4,4)$ sigma models proved to be
a direct harmonic counterpart of the nonlinear gauge theory action
constructed in \cite{{Ikeda},{austr}} by entirely
different reasoning! We believe that this exciting analogy
is a clue to the understanding of the intrinsic geometry of
general $(4,4)$ sigma models with torsion.

To avoid a possible confusion, it is worth mentioning that the theory
considered {\it is not} a supersymmetric extension of any
genuine $2D$ gauge theory:
there are no gauge fields in the multiplet of physical fields.
The only role of gauge invariance (\ref{gaugenab}) seems to consist in
ensuring the correct number of the sigma model physical fields
($4n$ bosonic and $8n$ fermionic ones).

It should be pointed out that it is the presence of the
antisymmetric potential $h^{2,2\;[N,M]}$ that makes the considered case
nontrivial and, in particular, the gauge invariance (\ref{gaugenab})
nonabelian. If
$h^{2,2\;[N,M]}$ is vanishing, the
invariance gets abelian and the constraints (\ref{2}) - (\ref{4})
are identically satisfied, while (\ref{1}) is solved by
\be \label{hzero}
h^{1,3\;M} = \nabla^{0,2}\Sigma^{1,1\;M}(q,u,v),\;\;
h^{3,1\;M} = \nabla^{2,0}\Sigma^{1,1\;M}(q,u,v)\;,
\ee
with $\Sigma^{1,1\;M}$ being an unconstrained prepotential. Then, using the
target space
gauge symmetry (\ref{hrep}), one may entirely gauge away $h^{1,3\;M},
h^{3,1\;M}$, thereby reducing (\ref{haction}) to the dual
action of twisted $(4,4)$ multiplets (\ref{dualq}). In the case of
one triple $q^{1,1}, \omega^{1,-1}, \omega^{-1,1}$ the potential
$h^{2,2\;[N,M]}$ vanishes identically, so the general action  (\ref{genact})
for $n=1$ is actually equivalent to (\ref{dualq}). Thus only
for $n\geq 2$ a new
class of torsionful $(4,4)$ sigma models comes out.
It is easy to see that the action (\ref{haction}) with non-zero
$h^{2,2\;[N,M]}$ {\it does not} admit any duality transformation to the form
with the superfields $q^{1,1\;M}$ only, because it is impossible to
remove the dependence on $\omega^{1,-1\;N}, \omega^{-1,1\;N}$ from the
equations for $q^{1,1\;M}$ by any
local field redefinition with preserving harmonic analyticity. Moreover,
in contradistinction to the constraints (\ref{qconstr}), these
equations are compatible
only with using the equation for $\omega$'s. So,
the obtained system definitely does not admit in general any dual
description in terms of twisted $(4,4)$ superfields. Hence, the
left and right complex structures on the target space can be non-commuting.
In the next section we will explicitly show this non-commutativity for
a particular class of the models in question.

\vspace{0.3cm}
\noindent{\bf 3. Harmonic Yang-Mills sigma models.}
Here we present a particular solution to the constraints (\ref{1})-(\ref{4}).
We believe that it shares many features of the general solution which
is as yet unknown.

It is given by the following ansatz
\bea
h^{1,3\;N} &=& h^{3,1\;N} \;=\; 0\;; \; h^{2,2} \;=\; h^{2,2}(t,u,v)\;, \;
\;t^{2,2} \;=\; q^{1,1\;M}q^{1,1\;M}\;; \nonumber \\
h^{2,2\;[N,M]} &=& b^{1,1} f^{NML} q^{1,1\;L}\;, \;b^{1,1} \;=\;
b^{ia}u^1_iv^1_a\;, \; b^{ia} = \mbox{const}\;,
\label{solut}
\eea
where the real constants $f^{NML}$ are totally antisymmetric. The constraints
(\ref{1}) - (\ref{3}) are identically satisfied with this
ansatz, while (\ref{4}) is now none other than the Jacobi identity
which tells us that the constants
$f^{NML}$ are structure constants of some real semi-simple Lie
algebra (the minimal possibility is $n=3$, the
corresponding algebra being $so(3)$). Thus the $(4,4)$ sigma models
associated with
the above solution can be interpreted as a kind of Yang-Mills theories in
the harmonic superspace. They provide the direct nonabelian generalization
of the twisted multiplet sigma models with the action (\ref{dualq}) which
are thus analogs of two-dimensional abelian gauge theory.
The action (\ref{haction}), related equations of motion and
the gauge transformation laws (\ref{gaugenab}) specialized to the
case (\ref{solut}) are as follows
\bea
S^{YM}_{q,\omega} &=&
\int \mu^{-2,-2} \{\; q^{1,1\;M} (\; D^{0,2}\omega^{1,-1\;M} +
D^{2,0}\omega^{-1,1\;M} + b^{1,1} \;\omega^{-1,1\;L} \omega^{1,-1\;N}
f^{LNM}\; ) \nonumber \\
&&+ \;h^{2,2}(q,u,v) \} \label{haction0}
\eea
\bea
&& D^{2,0} \omega^{-1,1\;N} + D^{0,2} \omega^{1,-1\;N} +
b^{1,1}\;\omega^{-1,1\;S} \omega^{1,-1\;T} f^{STN} \;\equiv\; B^{1,1\;N}
\;=\; - \frac{\partial h^{2,2}}{\partial q^{1,1\;N}}\;, \nonumber \\
&& D^{2,0}q^{1,1\;M} + b^{1,1} \;\omega^{1,-1\;N} f^{NML}  q^{1,1\;L}
\;\equiv \; \Delta^{2,0}q^{1,1\;M}\;=\; 0
\nonumber \\
&& D^{0,2}q^{1,1\;M} - b^{1,1} \; \omega^{-1,1\;N}f^{NML}q^{1,1\;L}
\;\equiv \; \Delta^{0,2}q^{1,1\;M}\;=\; 0
\label{heqmo0} \\
&&\delta \omega^{1,-1\;M}  \;=\;
\Delta^{2,0}\sigma^{-1,-1\;M} \;, \; \delta \omega^{-1,1\;M}  \;=\;
- \Delta^{0,2}\sigma^{-1,-1\;M}\;, \nonumber \\
&& \delta q^{1,1\;M} \;=\;
b^{1,1} \;\sigma^{-1,-1\;N} f^{NML} q^{1,1\;L}\;.
\label{gaugenab0}
\eea

These formulas make the analogy with two-dimensional nonabelian
gauge theory almost literal, especially for
\be \label{free}
h^{2,2} = q^{1,1\;M} q^{1,1\;M}\;.
\ee
Under this choice
$$
q^{1,1\;N} = -{1\over 2}\; B^{1,1\;N}
$$
by first of eqs. (\ref{heqmo0}), then two remaining equations are
direct analogs of two-dimensional Yang-Mills equations
\be \label{litan}
\Delta^{2,0} B^{1,1\;N} = \Delta^{0,2} B^{1,1\;N} = 0\;,
\ee
and we recognize (\ref{haction0}) and (\ref{heqmo0}) as the harmonic
counterpart of the first
order formalism of two-dimensional Yang-Mills theory.
In the general case $q^{1,1\;M}$ is a nonlinear function of
$B^{1,1\;N}$, however for $B^{1,1\;N}$ one still has the same equations
(\ref{litan}).

Now it is a simple exercise to see that in checking the integrability
condition (\ref{comm}) one necessarily needs first of eqs.
(\ref{heqmo0}), while in the abelian,
twisted multiplet case this condition is
satisfied without any help from the equation obtained by varying
the action (\ref{dualq}) with respect to $q^{1,1\;N}$.
This property reflects the fact that
the class of $(4,4)$ sigma models we have found cannot be described
only in terms of twisted $(4,4)$ multiplets (of course, in general the
gauge group has the structure of a direct product with
abelian factors; the
relevant $q^{1,1}$'s satisfy the linear twisted multiplet constraints
(\ref{qconstr})).

An interesting specific feature of this ``harmonic Yang-Mills theory'' is
the presence of the doubly charged ``coupling constant'' $b^{1,1}$
in all formulas, which is necessary for the correct balance of the
harmonic $U(1)$ charges. Since $b^{1,1} = b^{ia}u^1_iv^1_a$,
we conclude that in the geometry of the considered class
of $(4,4)$ sigma models a very essential role is played by the quartet
constant $b^{ia}$.
When $b^{ia} \rightarrow 0$, the nonabelian structure contracts into the
abelian one and we reproduce the twisted multiplet action (\ref{dualq}).
We shall see soon that $b^{ia}$ measures the ``strength'' of
non-commutativity of the left and right complex structures.

Let us limit ourselves to the simplest case (\ref{free}) and compute the
relevant bosonic sigma model action and complex structures. We will
do this to the first order in physical bosonic fields, which will be
sufficient to show the non-commutativity of complex structures.

We first impose a kind of Wess-Zumino gauge with respect to
the local symmetry (\ref{gaugenab0}). We choose it so as to gauge
away from $\omega^{1,-1\;N}$ as many components as possible,
while keeping $\omega^{-1,1\;N}$ and $q^{1,1\;N}$ arbitrary.
The gauge-fixed form of $\omega^{1,-1\;N}$ is as follows
\be \label{gaugefix}
\omega^{1,-1\;N} (\zeta, u, v) = \theta^{1,0\;\underline{i}}
\;\nu^{0,-1\;N}_{\underline{i}}(\zeta_R,v) + \theta^{1,0}\theta^{1,0}
\;g^{0,-1\;iN}(\zeta_R,v)u^{-1}_i
\ee
with
$$
\{ \zeta_R \} \equiv \{ x^{++},x^{--}, \theta^{0,1\;\underline{a}} \}\;.
$$
Then we substitute (\ref{gaugefix}) into (\ref{haction0})
with $h^{2,2}$ given by (\ref{free}), integrate over $\theta$'s and $u$'s,
eliminate infinite tails of decoupling auxiliary fields and, after this
routine work, find the physical bosons part of the action (\ref{haction0})
as the following integral over $x$ and harmonics $v$
\be   \label{bosact}
S_{bos} = \int d^2 x [dv] \left( {i\over 2}\;g^{0,-1\;iM}(x,v)\;
\partial_{--} q^{0,1\;M}_i (x,v) \right)\;.
\ee
Here the fields $g$ and $q$ are subjected to the harmonic differential
equations
\bea
&& \partial^{0,2} g^{0,-1\;iM} - 2 (b^{ka}v^{1}_a)\;f^{MNL} q^{0,1\;iN}
g^{0,-1\;L}_k
\;=\; 4i \partial_{++} q^{0,1\;iM} \nonumber \\
&& \partial^{0,2} q^{0,1\;iM} - 2f^{MLN} (b^{ka}v^1_a)\;q^{0,1\;L}_k \;
q^{0,1\;iN}
\;=\; 0  \label{eq12}
\eea
and are related to the initial superfields as
$$
q^{1,1\;M}(\zeta,u,v)| = q^{0,1\;iM}(x,v)u^{1}_i + ... \;, \;\;\;
g^{0,-1\;iN}(\zeta_R, v)| = g^{0,-1\;iN}(x,v)\;,
$$
where $|$ means restriction to the $\theta$ independent parts.

To obtain the ultimate form of the action as an integral over
$x^{++}, x^{--}$, we should solve eqs. (\ref{eq12}), substitute the
solution
into (\ref{bosact}) and do the $v$ integration. Here we solve (\ref{eq12})
to the first
non-vanishing order in the physical bosonic  field $q^{ia\;M}(x)$
which appears as the
first component in the $v$ expansion of $q^{0,1\;iM}$
$$
q^{0,1\;iM}(x,v) = q^{ia\;M}(x)v^1_a + ... \;.
$$

Representing
(\ref{bosact}) as
\be \label{bosact1}
S_{bos} = \int d^2x  \left( G^{M\;L}_{ia\;kb} \partial_{++} q^{ia\;M}
\partial_{--} q^{kb\;L} + B^{M\;L}_{ia\;kb} \partial_{++} q^{ia\;M}
\partial_{--} q^{kb\;L} \right)
\ee
where the metric $G$ and the torsion potential $B$ are,
respectively, symmetric
and skew-symmetric with respect to the simultaneous interchange of
the left and right triples of  their indices, we find that to
the first order
\be \label{GB}
G^{M\;L}_{ia\;kb} = \delta^{ML}\epsilon_{ik} \epsilon_{ab} -
{2\over 3} \epsilon_{ik} f^{MLN} b_{l(a} q^{l\;N}_{b)}\;,
\;\;
B^{M\;L}_{ia\;kb}  =  {2\over 3} f^{MLN} [b_{(i a}q^N_{k)b} +
b_{(ib}q^{N}_{k)a}]\;.
\ee
Note that an asymmetry between the indices $ik$ and $ab$ in the metric
is an artefact of our choice of the WZ gauge in the form (\ref{gaugefix}).
One could choose another gauge so that a symmetry between
the above pairs of $SU(2)$ indices is restored. Metrics in different
gauges are related via the target space $q^{ia\;M}$ reparametrizations.

Finally, let us compute, to the first order in $q^{ia\;M}$,
the left and right complex structures associated with the
sigma models at hand. Following the well-known strategy \cite{{GHR},{DS}},
we need: (i) to partially go on shell by eliminating the auxiliary
fermionic
fields; (ii) to divide four supersymmetries in every light-cone
sector into
a $N=1$ one realized linearly and a triplet of
nonlinearly realized extra supersymmetries; (iii) to consider the
transformation laws of the physical bosonic fields $q^{ia\;M}$ under
extra supersymmetries. The complex structures can be read off from these
transformation laws.

In our case at the step (i) we should solve some harmonic differential
equations of motion to express an infinite tail of auxiliary fermionic
fields
in terms of the physical ones and the bosonic fields $q^{ia\;M}$.
The step (ii) amounts
to the decomposition of the $(4,0)$ and $(0,4)$ supersymmetry parameters
$\varepsilon^{i\underline{i}}_{-}$ and $\varepsilon^{a\underline{a}}_+$
as
$$
\varepsilon^{i\underline{i}\;+}\equiv \epsilon^{i\underline{i}}
\varepsilon^+ + i \varepsilon^{(i\underline{i})\;+} \;, \;\;
\varepsilon^{a\underline{a}\;-} \equiv \epsilon^{a\underline{a}}
\varepsilon^- + i \varepsilon^{(a\underline{a})\;-}\;,
$$
where we have kept a manifest symmetry only with respect to the diagonal
$SU(2)$ groups
in the full left and right automorphism groups $SO(4)_L$ and $SO(4)_R$.
At the step (iii) we should redefine the physical fermionic fields so
that the singlet supersymmetries with the
parameters $\varepsilon_-$ and $\varepsilon_+$ be realized linearly.
We skip the details and present the final form of the on-shell
supersymmetry transformations of $q^{ia\;M}(x)$
\be
\delta q^{ia\;M} =
\varepsilon^+ \;\psi^{ia\;M}_+ +i \varepsilon^{(kj)\;+}
\;\left( F_{(kj)} \right)^{ia\;M}_{lb\;L}\;\psi^{lb\;L}_+
+
\varepsilon^- \;\chi^{ia\;M}_- + i\varepsilon^{(cd)\;-}
\;\left( \hat{F}_{(cd)} \right)^{ia\;M}_{lb\;L}\;\chi^{lb\;L}_-\;.
\ee
Introducing the matrices
$$
F^n_{(+)} \equiv  (\tau^n)^k_j F^{(j}_{\;\;\;\;k)}\;,\;\;
F^m_{(-)} \equiv  (\tau^m)^c_d \hat{F}^{(d}_{\;\;\;\;c)}\;,
$$
$\tau^n$ being Pauli matrices, we find that in the first order in
$q^{ia\;M}$ and $b^{ia}$
\bea
F^n_{(+)} &=& -i \tau^n \otimes I \otimes I + {i\over 3}\; [\;M_{(+)},
\tau^n \otimes I \otimes I\;]
\nonumber \\
F^n_{(-)} &=& -i I\otimes \tau^n \otimes I + {i\over 3}\;
[\;M_{(-)}, I\otimes \tau^n\otimes I\;] \label{cstr}
\eea
\be
\left( M_{(+)} \right)^{ia\;M}_{kb\;N} = -2\; f^{MLN}\left(
b^{(i}_{b} q^{a\;L}_{\;\;k)} + b^{(i a} q^{L}_{\;\;k)b} \right)\;, \;\;
\left( M_{(-)} \right)^{ia\;M}_{kb\;N} = 2\;f^{MLN} \;b^i_{(b}q^{a)\;L}_k
\;,\label{Matr}
\ee
where the matrix factors in the tensor products are arranged so that they
act, respectively,
on the subsets of indices $i,j,k,...$, $a,b,c,...$, $M,N,L,...$.

It is easy to check that the matrices $F^n_{(\pm)}$ to the first order
in $q$, $b$ possess all the standard properties of
complex structures \cite{GHR}. In particular, they form a
quaternionic algebra
$$
F^n_{(\pm)}F^m_{(\pm)} = - \delta^{nm} + \epsilon^{nms} F^s_{(\pm)}\;,
$$
and satisfy the covariant constancy conditions
$$
{\cal D}_{lc\;K} \left( F^n_{(\pm)} \right)^{ia\;M}_{kb\;N} =
\partial_ {lc\;K}\left( F^n_{(\pm)} \right)^{ia\;M}_{kb\;N} -
\Gamma^{\;\;\;\;\;\;\;\;\;\;\;jd\;T}_{(\pm)\;lc\;K\;\;kb\;N}
\left( F^n_{(\pm)} \right)^{ia\;M}_{jd\;T} +
\Gamma^{\;\;\;\;\;\;\;\;\;\;\;ia\;M}_{(\pm)\;lc\;K\;\;jd\;T}
\left( F^n_{(\pm)} \right)^{jd\;T}_{kb\;N} = 0
$$
with
$$
\Gamma^{\;\;\;\;\;\;\;\;\;\;\;jd\;T}_{(\pm)\;lc\;M\;\;kb\;N} \equiv
\Gamma^{\;\;\;jd\;T}_{lc\;M\;\;kb\;N} \mp
T^{\;\;\;jd\;T}_{lc\;M\;\;kb\;N}\;,
$$
where $\Gamma$ is the standard Riemann connection for the
metric (\ref{GB}) and $T$ is the torsion
$$
T_{ia\;M\;\;kb\;N\;\;ld\;T} = {1\over 2} \left( \partial_{ia\;M}
B^{N\;T}_{kb\;ld} + \;\;cyclic \right)\;.
$$
Of course, in the present case these properties are fulfilled
automatically as
we started with a manifestly $(4,4)$ supersymmetric off-shell superfield
formulation.

It remains to find the commutator of complex structures. The
straightforward computation (again, to the first order in fields)
yields
\bea
[\;F^n_{(+)}, F^m_{(-)}\;] &=&
(\tau^n\otimes I\otimes I) M_{(-)}
(I\otimes \tau^m\otimes I) +
(I\otimes \tau^m\otimes I) M_{(-)}
(\tau^n\otimes I\otimes I) \nonumber \\
&-& (\tau^n\otimes
\tau^m\otimes I)
M_{(-)} -
M_{(-)}(\tau^n\otimes
\tau^m\otimes I)
 \;\neq \; 0\;.
\label{commstr}
\eea

Thus in the present case in the bosonic sector we encounter a more general
geometry compared to the one discussed
in \cite{{GHR},{HoPa}}. The basic characteristic feature of this new
geometry
is the non-commutativity  of the left and right complex structures.
It is easy to check this property also for more general
potentials $h^{2,2}(q,u,v)$ in (\ref{haction0}). It
seems obvious that the general case (\ref{haction}),
({\ref{1}) - (\ref{4}) reveals the same feature.
Stress once more that this important property
is related in a puzzling way to
the nonabelian structure of the analytic superspace
actions (\ref{haction0}),
(\ref{haction}): the ``coupling constant'' $b^{1,1}$
(or the Poisson potential
$h^{2,2\;[M,N]}$ in the general case) measures the strength of the
non-commutativity of complex structures.

\vspace{0.3cm}
\noindent{\bf 4. Conclusion.}
To summarize, proceeding from an analogy with
the $SU(2)$ harmonic superspace description of $(4,4)$ hyper-K\"ahler
sigma models, we have constructed off-shell $SU(2)\times SU(2)$ harmonic
superspace actions for
a new wide class of $(4,4)$ sigma models with torsion and
non-commuting left and right complex structures on the bosonic target.
This non-commutativity is directly related to
the remarkable non-abelian Poisson gauge structure of these actions. One
of the most characteristic features of the general action is the
presence of an infinite number of auxiliary fields and the lacking of
dual-equivalent
formulations in terms of $(4,4)$ superfields with finite sets of auxiliary
fields. It would be interesting to see whether such formulations exist for
some particular cases, e.g., those corresponding to the bosonic
manifolds with
isometries. An example of $(4,4)$ sigma model with non-commuting
structures which admits such a formulation has been given in \cite{RIL}.

The obvious problems for further study are to compute the relevant
metrics and torsions in a closed form and to try to utilize the
corresponding
manifolds as backgrounds for
some superstrings. An interesting question is
as to whether the constraints
(\ref{1}) - (\ref{4}) admit solutions corresponding to the $(4,4)$
supersymmetric group
manifold WZNW sigma models. The list of appropriate group manifolds
has been given in \cite{belg}. The lowest dimension manifold
with non-commuting left and right structures \cite{RSS}
is that of $SU(3)$. Its dimension 8 coincides
with the minimal bosonic manifold dimension at which a non-trivial
$h^{2,2\;[M,N]}$ in
(\ref{haction}) can appear.

Of course, it remains to prove that the
action (\ref{haction})
indeed describes most general $(4,4)$ models with torsion.
One way to do this is to start, like in the hyper-K\"ahler and
quaternionic cases \cite{{GIOSap},{quat}}, with the constrained formulation
of the relevant geometry in a
real $4n$ dimensional manifold and to reproduce
the potentials in (\ref{haction}) as some fundamental objects
which solve the initial constraints.  But even before performing such an
analysis, there arises the question of how general our original ansatz
(\ref{genact}) for the
action is. In fact, in \cite{Iv} we have
chosen a more general ansatz and shown that at least for
four-dimensional bosonic manifolds it is effectively
reduced to (\ref{genact}) and hence to (\ref{haction})
after the combined use of the target space diffeomorphisms and the
intregrability condition (\ref{comm}). For general manifolds of
the dimension $4n, \;n\geq 2$, this still needs to be checked, in
\cite{Iv} only some heuristic arguments in favour of this have
been adduced.
Another point is that the constrained superfield $q^{1,1\;M}$ the
dual action of which was a starting point of our construction,
actually
comprises only one type of $(4,4)$ twisted multiplet \cite{IK}.
There exist other
types which differ in the $SU(2)_L\times SU(2)_R$ assignment of
their components
\cite{{GHR},{GI}}. At present it is unclear how
to simultaneously describe all of them
in the framework of the $SU(2)\times SU(2)$ analytic
harmonic superspace. Perhaps, their actions are related to
those of $q^{1,1}$
by a kind of duality transformation. It may happen, however, that
for their self-consistent description one will need a more general
type
of $(4,4)$ harmonic superspace, with the whole $SO(4)_L\times SO(4)_R$
automorphism group of $(4,4)$ SUSY harmonized. The relevant actions
will be certainly more
general than those considered in this paper.

\vspace{0.3cm}
\noindent{\bf Acknowledgements.}
The author thanks ENSLAPP, ENS-Lyon, where this work has been
finished, for kind hospitality and Francois Delduc for interest in
the work and useful discussions.  A partial support from
the Russian Foundation of Fundamental
Research, grant 93-02-03821, and the International Science Foundation,
grant M9T000, is acknowledged.

\end{document}